\documentclass[10pt,conference]{IEEEtran}
\IEEEoverridecommandlockouts

\usepackage{hyperref}
\hypersetup{
    colorlinks=true,
    linkcolor=blue,
    urlcolor=cyan,
    citecolor=blue,
}

\usepackage{balance}
\usepackage{amsmath}
\usepackage{amssymb}
\usepackage{amsfonts}
\usepackage{algorithmic}
\usepackage{graphicx}
\usepackage{textcomp}
\usepackage{xcolor}
\usepackage{algorithmic}
\usepackage{graphicx}
\usepackage{textcomp}
\usepackage{xcolor}
\usepackage{verbatim}
\usepackage{listings}
\usepackage{multirow}
\usepackage{enumitem}
\usepackage{framed}
\usepackage{colortbl}
\usepackage{soul}
\usepackage{booktabs} 
\usepackage{tcolorbox}

\usepackage[backend=biber,style=ieee,natbib=true,maxbibnames=9,maxcitenames=2,mincitenames=1]{biblatex} 
\definecolor{lightgray}{gray}{0.95}
\definecolor{darkblue}{rgb}{0,0,0.5}
\definecolor{lightred}{rgb}{1, 0.8, 0.8}
\definecolor{lightyellow}{rgb}{1, 1, 0.8}
\definecolor{lightgreen}{rgb}{0.6, 1, 0.6}
\definecolor{mint}{rgb}{0.74, 0.99, 0.79}
\definecolor{forestgreen}{rgb}{0.13, 0.55, 0.13}
\definecolor{goldenrod}{RGB}{255, 185, 65}

\def\BibTeX{{\rm B\kern-.05em{\sc i\kern-.025em b}\kern-.08em
    T\kern-.1667em\lower.7ex\hbox{E}\kern-.125emX}}

\begin{document}

\title{Ensuring Open Source Integrity: The Intersection of Copy-Based Reuse and License Compliance
\thanks{Replication package available at: \url{https://zenodo.org/records/14061115}}
}

\author{
\IEEEauthorblockN{1\textsuperscript{st} Mahmoud Jahanshahi}
\IEEEauthorblockA{
\textit{University of Tennessee}\\
Knoxville, USA \\
mjahan@utk.edu}
\and
\IEEEauthorblockN{2\textsuperscript{nd} Bogdan Vasilescu}
\IEEEauthorblockA{
\textit{Carnegie Mellon University}\\
Pittsburgh, USA \\
vasilescu@cmu.edu}
\and
\IEEEauthorblockN{3\textsuperscript{rd} Audris Mockus}
\IEEEauthorblockA{
\textit{University of Tennessee}\\
Knoxville, USA \\
audris@utk.edu}
}

\maketitle

\begin{abstract}
  As other creative work, source code is protected by
  copyright.
  The owner can license the work, e.g., to permit copy and other kinds of
  use, and even 
  start legal proceeding against license violators.
  However, source code can be reused in subtle ways, e.g., 
  via copying without explicit package manager dependencies,
  making it hard to reason about potential license noncompliance.  
  Using the World of Code infrastructure approximating 
  the entirely of open source software, in this paper we create a 
  copy-based code reuse network mapping direct copying across projects, 
  and use it to quantify the extent of potential license noncompliance 
  across the entire open source ecosystem.
  In addition, we estimate regression models to understand whether code 
  copying is affected by the origin project's license, and, if so, 
  how it varies with other project characteristics.

  \looseness=-1
  We find that code in repositories with permissive licenses, such as MIT and
  Apache, shows higher likelihood of reuse across programming
  languages. In contrast, copyleft licenses, like the GPL, exhibit mixed effects.
  Public domain licenses, despite their aim of allowing unrestricted
  use, are associated with lower likelihood of copy-based reuse. 
  A widespread potential license
  noncompliance appears to accompany copy-based reuse, with 39.4\%
  of project combinations at potential noncompliance risk,
  particularly when licenses are unclear or absent. 
  Our findings reveal that only 2.43\% of reuse detected through 
  the copy-based network was discoverable via dependency analysis, 
  highlighting the limitations of existing dependency-tracking tools 
  in capturing copy-based reuse. This gap underscores the need for 
  more advanced methods to ensure license compliance in open source 
  projects, from nudging developers to set appropriate license 
  templates to flagging potential noncompliance due to license 
  changes across copy origin and destination projects.
\end{abstract}


\section{Introduction}\label{intro}

Open Source Software (OSS) plays a critical role in software development and distribution across various industries.
A fundamental aspect of OSS is its licensing, which dictates how software can be reused, modified, and redistributed.

However, not all OSS 
participants are aware of or follow the ramifications of 
licensing~\cite{almeida2017software}. For example, 
their code is under exclusive copyright by default if no license is specified. 
The lack of knowledge, and/or, possibly, the lack of enforcement,
results in reuse of unlicensed code in contexts where license
incompatibilities are likely. For example, with the rise of 
large language models (LLMs), a significant portion of the 
massive source code used for training consists of code without 
any license~\cite{yu2023codeipprompt,xu2024first,jahanshahi2025cracks}.

OSS licenses are typically categorized into permissive licenses (e.g., MIT, Apache), copyleft licenses (e.g., GPL), weak copyleft licenses (e.g., LGPL), public domain licenses, and others with specific conditions (e.g., Creative Commons).
Each type of license imposes distinct obligations on developers and users, making the choice of license a pivotal factor in determining the extent and manner in which a project's code can be reused.
Moreover, creative works, such as code, are protected by copyright by default if no license is specified.
Despite these legal restrictions, code without a license is often copied in practice~\cite{vendome2018distribute} and even used to train Large Language Models~\cite{xu2024first}.  

This study aims to enhance understanding of the extent to which, and the contexts in which, different OSS license types affect software reuse in copy-based reuse networks, where artifacts are copied from one repository to another.

While prior research has primarily focused on dependency-based reuse, where projects formally declare dependencies on external libraries, copy-based reuse---where code is directly copied between projects---introduces unique challenges regarding license compliance and tracking, because there is typically no trace of the copying.
Studies highlight that identifying the exact origin of reused OSS components remains a significant challenge, underscoring the need for more effective tools to track code provenance, particularly to ensure compliance with copyleft licenses~\cite{tuunanen2021tool}.

Although copy-based reuse is common in OSS development~\cite{jahanshahi2024beyond}, it is often overlooked in studies that focus exclusively on dependencies managed through package managers~\cite{fendt2019open,phipps2020continuous,german2010understanding}.
While the decision to copy an artifact from an upstream project may be driven by factors largely unrelated to license compatibility, the type of license should still play a significant role, particularly if the license of the copied artifact is ultimately incompatible with that of the reusing project.

Specifically, we answer the following research questions:
\begin{itemize}
  \item \textbf{RQ1}: How does the license type of the upstream
      project affect the probability of its artifacts getting copied?
  \item \textbf{RQ2}: How widespread is potential license
      noncompliance in copy-based reuse network? 
\end{itemize}

We begin by reviewing the literature on code copying to identify key 
factors driving this phenomenon. 
Next, we use the World of Code (WoC) infrastructure, which 
provides comprehensive, cross-referenced data on the global OSS ecosystem, to operationalize these factors and create a curated
dataset of copying instances, including the licenses of both upstream and
downstream projects. Finally, we fit a model that examines the probability 
of a project's artifacts being reused based on its license, 
while controlling for other contextual factors.

Our findings indicate that permissive licenses, such as MIT and
Apache, are consistently associated with higher reuse rates across multiple
programming languages. 
In contrast, copyleft licenses, like GPL, display more complex reuse patterns.
While they are associated with higher rates of reuse in certain cases, such as in JavaScript projects, they are generally associated with lower reuse when
factors like project size and activity are considered. 
Interestingly, projects under public domain licenses, which are intended to permit unrestricted reuse, tend to experience lower reuse rates.
This suggests that legal uncertainties surrounding these licenses may deter developers from reusing the code.

One notable issue we uncovered is the prevalence of license noncompliance in copy-based reuse, especially when projects either lack a clear license or use incompatible licenses, posing legal risks for developers and organizations alike.
License noncompliance in software reuse is not just a theoretical concern but has resulted in significant legal disputes in the software industry.
A notable example is the \textit{Jacobsen v. Katzer} case~\cite{shagall2008jacobsen},
wherein the court upheld the enforceability of open source licenses under copyright law.
Jacobsen, the creator of the Java Model Railroad Interface (JMRI) project, sued Katzer for incorporating JMRI's code into commercial software without adhering to the terms of the project's Artistic License.
The court's decision affirmed that violating open source license terms constitutes copyright infringement, emphasizing the legal obligations developers have when reusing code.

Another case illustrating the repercussions of license noncompliance involves the GPL-licensed \textit{BusyBox} OSS project~\cite{sflc2007busybox}.
\textit{BusyBox} developers filed lawsuits against several companies for distributing their software within commercial products without complying with GPL terms.
These companies failed to provide access to the source code and did not include the GPL license text with their products, both required under the GPL.
The legal actions often resulted in settlements where the offending companies agreed to release the source code and comply with the GPL terms.

These real-world examples underscore the importance of understanding and adhering to license terms, especially in copy-based reuse where code is directly replicated between projects.
Noncompliance not only exposes developers and organizations to legal risks but also undermines the collaborative ethos of the OSS community~\cite{german2010understanding}.
It can deter developers from contributing or reusing code due to fears of infringement, thereby stifling innovation and collaboration.
Therefore, ensuring proper license compliance is essential for fostering trust and sustainability in open source software development.

Finally, our study reveals that traditional tools focused on dependency tracking fail to capture a substantial number of reuse cases occurring through direct code copying.
This highlights the need for more sophisticated tools capable of detecting direct code copying at scale, to improve license compliance monitoring within the OSS ecosystem.

\section{Related Work and Knowledge Gaps}
\label{sec:relwork}

\subsection{Software Reuse} 

In open source software, the reuse within supply chains can be categorized based on how open source components are integrated and used in software projects~\cite{mockus2019insights,isectut22,ase23}.

\subsubsection{Dependency-Based Reuse} This category involves incorporating open source libraries and packages as dependencies in a project.
Package managers like NPM for JavaScript, pip for Python, or Maven for Java are typically used to manage these dependencies.
If not properly overseen, reliance on these dependencies can introduce vulnerabilities and risks~\cite{yan2021estimating}.

\subsubsection{Copy-Based Reuse (Our Focus)} 
In copy-based reuse, developers directly copy code from OSS projects, e.g., a utility function~\cite{jahanshahi2024beyond}, into their own projects.
While this approach is quick, it can lead to challenges in maintaining and updating the copied code.
Therefore, it's essential to track and manage these copies to ensure they remain secure and up-to-date~\cite{ladisa2023sok}.

Previous studies have identified several factors that influence the likelihood of a project's artifacts being reused through copy-based methods~\cite{jahanshahi2024beyond}.
One key factor is \textbf{project activity}, typically measured by the number of commits.
Projects with a higher commit count are generally more active and frequently updated, making them attractive to developers seeking reliable and current code snippets~\cite{koch2002effort}.
Another important factor is \textbf{project size}, often indicated by the number of files.
Larger projects tend to offer a broader range of functionalities and code examples, increasing the likelihood that other developers will find useful code for reuse.
This extensive codebase provides a valuable resource for copy-based reuse~\cite{mockus2007large}.
The collaborative nature of a project also plays a role.
Metrics such as the \textbf{number of authors} 
reflect the volume and diversity of expertise within a project’s contributor base.
Projects with more contributors tend to benefit from enhanced innovation and decentralized communication, which can improve the development process~\cite{crowston2005social} and increase the likelihood of reuse~\cite{jahanshahi2024beyond}.
Community engagement and popularity, often approximated by metrics such as the \textbf{number of forks and stars} on platforms like GitHub, further explain reuse potential~\cite{tsay2014influence, borges2016predicting}.
Projects with more forks and stars are more visible and reputable within the developer community, increasing trust and making their code more likely to be reused~\cite{jahanshahi2024beyond}.
These indicators reflect community interest and endorsement, enhancing the project's appeal as a resource.

The \textbf{maturity and stability} of a project, assessed through its duration of activity, age, and activity fluctuations (burstiness), also correlate with its reuse potential~\cite{jahanshahi2024beyond}.
Mature projects with sustained activity over a long period are often viewed as stable and reliable.
Consistent development without erratic bursts signals a well-maintained project, increasing the likelihood that its code will be reused~\cite{gamalielsson2014sustainability}.
Additionally, a project's community culture and technical characteristics---approximated by its \textbf{primary programming language}---play a significant role in explaining its reuse potential~\cite{jahanshahi2024beyond}. Different programming languages vary in popularity, community support, and ecosystem maturity~\cite{bissyande2013popularity}.
Projects written in widely adopted languages such as Python, JavaScript, or Java are more accessible to a larger pool of developers, thus increasing the chances of their code being reused.
Moreover, the programming language reflects the community’s coding conventions, documentation practices, and collaboration norms, which can make the project more appealing for developers looking to incorporate its code into their own work.

Finally, the literature highlights that \textbf{permissive licenses}, such as MIT and BSD, are generally associated with higher reuse rates compared to restrictive licenses like GPL~\cite{kashima2011investigation,brewer2012effects}.
Additionally, a \textbf{delay in license adoption} for a project might increase the chances of its artifacts being reused as the absence of a clear license can create ambiguity, leading developers to assume permissibility, thus fostering reuse even if unintended by the project maintainers.
However, these conclusions are based on simple statistical analyses that do not account for the critical factors influencing reuse discussed earlier.
Therefore, it is possible that the observed effect of licensing on reuse is not as strong as suggested, or that other variables may be driving these patterns.
A more comprehensive analysis---one that controls for these additional variables---is necessary to determine whether licensing independently influences reuse or if the previously-reported results are mostly shaped by other project characteristics.
Towards answering \textbf{RQ1}, we posit two concrete hypotheses:
\begin{itemize}
    \item \textbf{Hypothesis (H1a)}: Projects using permissive licenses, when controlling for other context factors, have a higher likelihood of their artifacts being reused via copying.
    \item \textbf{Hypothesis (H1b)}: Projects using restrictive licenses, when controlling for other context factors, have a lower likelihood of their artifacts being reused via copying.
\end{itemize}

\subsection{Open Source Licenses}

There are many licenses for open source code, each with its own requirements and restrictions.

\textbf{Permissive licenses}, such as MIT and Apache-2.0, typically allow for extensive reuse with few restrictions.
They usually require only attribution and permit integration with other license types, offering significant flexibility~\cite{laurent2004understanding}.
In contrast, \textbf{copyleft licenses}, such as the GPL, require that any derivative work be distributed under the same license.
Noncompliance can occur if copyleft-licensed code is combined with code under a non-copyleft license without adhering to the copyleft terms.
For example, incorporating GPL-licensed code into proprietary software without releasing the combined code under the GPL would violate the license~\cite{stallman2002free}.
This principle ensures that all modifications and derivative works remain free, preserving software freedom~\cite{lessig2004big}.
\textbf{Weak copyleft licenses}, such as the LGPL, are less restrictive than full copyleft licenses.
They permit linking with proprietary software without requiring the entire work to be open sourced, as long as the LGPL-covered components remain modifiable and separable.
However, it's important to carefully consider the terms to avoid violations, particularly regarding modification and distribution~\cite{rosen2005open}.
\textbf{Conditional open licenses}, including many Creative Commons licenses, offer specific conditions for use.
For example, CC-BY licenses require attribution, while CC-BY-SA licenses require derivative works to be licensed under the same terms.
These licenses can include share-alike clauses, which impact how code can be distributed, especially if combined with other licenses with different terms.
While these licenses are more commonly used for creative works than software, they can still impact code distribution.
\textbf{Public domain} and \textbf{license-free} software code generally impose no restrictions on reuse, as they are not protected by copyright.
Works in the public domain can be freely used, modified, and distributed.
Finally, projects with \textbf{no explicit license} (not to be confused with license-free) present significant legal risks.
By default, all rights are reserved under copyright law, meaning that reuse, modification, or distribution may be restricted without the author's explicit permission~\cite{valimaki2005rise}.
This lack of clarity can lead to potential legal issues, as the permissions for using the software are not clearly defined.

\subsection{Open Source License Compliance}
License compatibility is a critical concern in OSS development.
Projects often encounter significant difficulties when integrating components with conflicting licenses~\cite{di2010exploratory}.
Ensuring compliance with open source licenses is also a major concern for companies incorporating OSS into their products.
\citet{german2010understanding} emphasized the need for auditing OSS distributions to ensure adherence to license terms, especially in scenarios where components with varying licenses are integrated.
\citet{wu2024large} conducted a large-scale empirical analysis on the usage of open source licenses, highlighting the practices and challenges developers face.
Their findings revealed 
frequent misunderstandings and misapplications of licenses, especially in large-scale projects.
\citet{cui2023empirical} created a tool called DIKE to detect license conflicts in over 16,000 popular free and OSS software, 
finding that over 25\% had conflicts.
In addition, their study suggests that these conflicts often arise from misinterpretations of license terms and the challenges of handling multi-license environments.
Finally, \citet{mathur2012empirical} conducted an empirical study on license violations resulting from code reuse across 1,423 projects, uncovering numerous instances of license incompatibilities.

In addition, many developers involved in OSS projects do not fully understand the implications of the licenses they use.
\citet{almeida2019investigating,almeida2017software} revealed gaps in developers’ knowledge of licensing issues, which can result in non-compliance, particularly in complex projects that integrate multiple OSS components.
Moraes et al.~\cite{moraes2021one} and Qiu et al.~\cite{qiu2021empirical} focused on the JavaScript ecosystem, investigating the effects of multi-licensing and license violations related to dependencies.
Their findings show that the complex network of dependencies in JavaScript projects frequently results in unintentional license violations, highlighting the need for improved dependency management practices.
Feng et al.~\cite{feng2019open} investigated license violations in large-scale binary software, 
revealing that many projects unintentionally breach license terms due to the complexities involved in binary distribution.
Finally, \citet{papoutsoglou2022analysis} examined licensing questions on Stack Exchange sites, 
their results 
showing that many developers find it challenging to grasp licensing terms, leading to frequent inquiries about compliance and compatibility issues.

Studies have also demonstrated that a project's declared license is not always reliable~\cite{german2010understanding, reid2023applying, wolter2023open}.
For example, in a study of OSS projects on GitHub, \citet{wolter2023open} 

discovered that in approximately 50\% of the projects analyzed, the top-level declared license did not fully reflect all the licenses present within the project, 
emphasizing the importance of improved education and automated tools for ensuring compliance.
Moreover, \citet{wu2015method} found instances where the license of a source code file was altered after being copied, both by the original author of the code, and by the reuser; the latter likely constitute a license violation.

The complexities of OSS licensing are further heightened by the widespread practice of copy-based code reuse, which can lead to unintended license violations~\cite{jahanshahi2024beyond}.
Managing license compliance in these scenarios is crucial for maintaining the integrity of open source projects.
\citet{jahanshahi2024beyond} showed that 80\% of OSS projects have practiced copy-based reuse, including large and popular projects.
They also demonstrated that a significant portion of the reused artifacts originate from small, lesser-known projects.
Given the widespread prevalence of copy-based reuse and the complexities of tracking the origins of artifacts, we anticipate a high potential risk of license noncompliance in this type of reuse.
This issue becomes even more critical considering that copy-based reuse is generally overlooked both by prior research and practitioners, thereby increasing the overall risk for the OSS community.
Towards answering RQ2, we hypothesize that:
\begin{itemize}
    \item \textbf{Hypothesis (H2a)}: Copy-based reuse carries a high risk of license noncompliance due to compounded complexities in tracking artifact origins.
    \item \textbf{Hypothesis (H2b)}: By overlooking copy-based code reuse, we are missing a significant portion of license noncompliance issues in open source software.
\end{itemize}

\subsection{Our Study vs Prior Work}

Our work offers a comprehensive and practical approach to identifying and addressing potential licensing issues arising from copy-based reuse in open source software, and it distinguishes itself from prior research in several ways:

\subsubsection{Comprehensive Identification of Licenses} 

Most studies, including those by Wu et al.~\cite{wu2024large} and Xu et al.~\cite{xu2023lidetector}, rely heavily on explicit license declarations in metadata files.
Others, like Feng et al.~\cite{feng2019open}, use static analysis of binaries to detect embedded license texts.
However, these approaches can miss licenses that are not explicitly declared or are located in less conventional directories.
In contrast, our work analyzes a comprehensive dataset~\cite{jahanshahi2024license} created by
exhaustively scanning the entire OSS landscape (as reflected in the World of Code~\cite{ma2019world}) for files containing the word ``license'' in their filepath.
This includes not only standard license files but also any file that may contain licensing information, ensuring no (obvious) potential license data is overlooked.

\subsubsection{Scale and Scope of Analysis} 

Previous studies often concentrate on specific platforms (e.g., GitHub), particular package manager ecosystems (e.g., NPM), or a narrow range of licenses (e.g., OSI-approved), leading to a partial approach to license detection and analysis.
For instance, the work by \citet{feng2019open} maps binary code to source code, detecting instances where code is directly incorporated into binary software.
While theoretically feasible, this approach encounters significant scalability challenges due to the substantial processing power required for large-scale analysis.
The computational demands of binary-to-source mapping render it impractical for use across the entire open-source ecosystem, especially when dealing with diverse binaries and platforms.
In contrast, our work examines the entire open-source ecosystem, offering a more comprehensive, cross-platform perspective on licensing violations.
By focusing on scalable methods that encompass various licenses, package managers, and code reuse practices, our approach addresses the scale limitations of prior studies, while providing a more practical solution for detecting license violations across the open-source landscape.
Moreover, our approach is not limited to code reuse; 
it can identify reuse across various types of artifacts, including documentation, configuration files, and other non-code components.
This capability offers a more comprehensive perspective on reuse and the associated licensing challenges.

\subsubsection{Controlling for Project Context}

Compared to prior research, our work reflects a more nuanced analysis of the relationship between software licensing and code reuse.
Unlike earlier studies that primarily used bivariate statistical correlations~\cite{kashima2011investigation,brewer2012effects}, we use a more sophisticated methodology that accounts for covariates, such as project size, community activity, and programming language.
By controlling for these factors, our work provides a clearer understanding of whether licensing type---permissive versus restrictive---independently influences reuse probability.
This allows us to re-examine the claims made in prior studies and offers more robust insights into the impact of licensing on OSS reuse.

\subsubsection{Analysis of License Violations in Copy-based Reuse} 

While many studies have explored license conflicts, few have employed a copy-based reuse network approach to understand the reuse patterns and potential violations and they often focus only on dependency-based reuse networks. 
As shown recently~\cite{jahanshahi2024beyond}, copy-based reuse is prevalent and contributes significantly to reuse practices in OSS.
Our research uses the copy-based reuse network to identify potential license violations due to license incompatibilities and reuse patterns, providing a novel perspective on how licenses interact across repositories.
This not only reveals license conflicts but also traces their origins, facilitating targeted resolutions and ensuring compliance across the software ecosystem.

\section{Methodology}\label{method}
\subsection{World of Code Infrastructure}

World of Code (WoC)\footnote{https://worldofcode.org}~\cite{ma2019world} is an infrastructure developed to cross-reference source code change data across the entire OSS community, enabling sampling, measurement, and analysis both within and across software ecosystems~\cite{ma2019world,ma2021world}.
Essentially, WoC functions as a software analysis pipeline, handling data discovery and retrieval, storage and updates, as well as the transformations and augmentations required for subsequent analytical tasks~\cite{ma2021world}.

WoC provides various maps that link git objects and metadata (e.g., commits, blobs, authors) to each other. 
It also offers more advanced maps, such as project-to-data connections (e.g., project-to-author), author aliasing~\cite{fry2020dataset}, and project deforking maps~\cite{mockus2020complete}.
In our study, we use WoC’s project-to-license (P2L) map~\cite{jahanshahi2024license}, which shows the licenses committed to each project in its most recent state (Version V of WoC, updated in March 2024).
\footnote{Version V, the most recent at the time of this study.}
Additionally, we apply the concept of deforked projects, as introduced by \citet{mockus2020complete}, to minimize potential biases caused by forks and duplicates of the same project.
Throughout this paper, the term ``project'' refers to these deforked projects unless stated otherwise.

\subsection{Copy-based Reuse Network}
In the context of OSS development, analyzing code reuse is essential for understanding the propagation of software components and the associated licensing implications.
Traditionally, the literature has primarily focused on dependency-based reuse, where the relationships between projects are analyzed based on declared package-manager dependencies, such as libraries or frameworks included in a project.
While dependency-based analysis provides valuable insights into how projects rely on external components, it often overlooks the more granular aspect of direct code copying, which can occur independently of formal dependencies.
Such practices are common in OSS projects but often remain undetected in dependency-based analyses, as shown by \citet{jahanshahi2024beyond}.
By mapping these direct copies, a copy-based reuse network provides a comprehensive view of code propagation, highlighting the actual flow of code between projects.

In the realm of license compliance, dependency-based analysis often focuses on the licenses of declared dependencies.
However, license obligations are not limited to these formal dependencies.
Copy-based reuse, particularly when undetected, can lead to unintentional license violations.
By mapping direct code copying, a copy-based reuse network allows for the identification of potential licensing conflicts that may arise from incorporating code with incompatible license terms, when the code wasn't part of a declared dependency.

To track this kind of reuse, WoC offers the Ptb2Pt map, which lists reused blobs (i.e., file versions) along with the creator, reuser, and the time each project first committed that blob~\cite{jahanshahi2024dataset}.
This map is created by sorting the timestamps of all commits creating a blob, with the project associated with the earliest commit identified as the creator.
Projects with any subsequent commits are then identified as reusers of that blob.

Next, since we are interested in a project-level analysis and since projects may reuse many blobs from one another, we further
aggregated the data based on unique combinations of upstream and downstream projects, counting the number of reused blobs between these projects for each combination.
The total number of unique upstream-downstream project combinations was 1,815,996,757.
Given our focus on potential license noncompliance, 
we excluded all instances of code reuse where the same entity (account) owns both the source and target projects.
This further reduced the data down to 1,788,541,220 combinations, indicating that about 1.5\% of reuse instances occurred between projects with the same owner.

Furthermore, given that the distribution of copied blob counts between projects is heavily right-skewed, we analyze potential noncompliance within the reuse network in two distinct modes to gain better insights. 
First, we consider \textbf{complete reuse}, including any instance where \textbf{at least one blob} has been copied in our analysis.  
Second, we refine the data to focus on reuse instances where \textbf{at least ten blobs} have been copied between upstream and downstream projects, as a proxy for more deliberate and \textbf{substantial reuse}.

\subsection{Potential License Noncompliance}

Noncompliance can manifest in various ways, often resulting in substantial legal and operational risks.
For example, it can occur when there are conflicts or misunderstandings regarding the terms and conditions of these licenses.
To better understand the associated risks, we categorize the outcomes of license combinations into three levels: \textit{No Issues}, \textit{Potential Issue - Low Risk}, and \textit{Potential Issue - High Risk}.

\paragraph{No Issues} This category covers situations where combining different licenses does not create legal or practical issues.
Projects under these licenses can be freely integrated, modified, and redistributed without concern for restrictive terms.
For instance, public domain and permissive licenses, such as the MIT or Apache 2.0 licenses, generally impose minimal restrictions.
These licenses are designed to encourage widespread use and modification, making them highly compatible with other licenses.
Their permissive nature ensures that they do not impose additional restrictions on combined works, allowing for seamless integration with other projects~\cite{laurent2004understanding,rosen2005open}.

\paragraph{Potential Issue - Low Risk} These combinations produce minor or manageable incompatibilities, such as
attribution, notice preservation, or compliance with specific conditions. 
For example, weak copyleft licenses, such as the LGPL, allow linking with proprietary software, provided that modifications to the LGPL-covered code remain open-source.
This flexibility reduces the likelihood of significant legal issues when combined with other licenses.
Similarly, licenses such as the Mozilla Public License (MPL) require modified files to be distributed under the same license but allow linking with other code, thus posing only minor issues~\cite{fitzgerald2006transformation}.

\paragraph{Potential Issue - High Risk} 
These combinations can create substantial legal or practical obstacles.
These issues typically arise from strict copyleft provisions or other incompatible conditions that limit the redistribution, modification, or integration of the software.
For instance, strong copyleft licenses, such as the GPL, require that any derivative works be licensed under the same terms.
This requirement can conflict with other licenses, especially those that are more permissive or do not allow for relicensing under the GPL's terms.
Such incompatibilities can prevent the distribution of combined works, necessitating careful consideration and potentially complex legal negotiations~\cite{stallman2002free,moglen2001free}.

The matrix in Table~\ref{tab:license_reuse} outlines various reuse scenarios and the corresponding risks of license noncompliance.

\begin{table}[t]
\centering
\caption{License Reuse Matrix and Potential Noncompliance Scenarios}
\resizebox{0.99\linewidth}{!}{
\begin{tabular}{l|cccccc}
    \toprule
    \multicolumn{1}{r|}{To} & \multirow{2}{*}{\textbf{Permissive}} & \multirow{2}{*}{\textbf{Copyleft}} & \textbf{Weak} & \textbf{Conditional} & \textbf{Public} & \textbf{No} \\
    From & & & \textbf{Copyleft} & \textbf{Open} & \textbf{Domain} & \textbf{License} \\
    \midrule
    \textbf{Permissive} & \textcolor{forestgreen}{No} & \textcolor{forestgreen}{No} & \textcolor{forestgreen}{No} & \textcolor{forestgreen}{No} & \textcolor{forestgreen}{No} & \textcolor{goldenrod}{Low} \\
    \textbf{Copyleft} & \textcolor{red}{High} & \textcolor{forestgreen}{No} & \textcolor{red}{High} & \textcolor{red}{High} & \textcolor{red}{High} & \textcolor{red}{High} \\
    \textbf{Weak Copyleft} & \textcolor{goldenrod}{Low} & \textcolor{forestgreen}{No} & \textcolor{forestgreen}{No} & \textcolor{goldenrod}{Low} & \textcolor{goldenrod}{Low} & \textcolor{red}{High} \\
    \textbf{Conditional} & \textcolor{goldenrod}{Low} & \textcolor{red}{High} & \textcolor{red}{High} & \textcolor{red}{High} & \textcolor{goldenrod}{Low} & \textcolor{red}{High} \\
    \textbf{Public Domain} & \textcolor{forestgreen}{No} & \textcolor{forestgreen}{No} & \textcolor{forestgreen}{No} & \textcolor{forestgreen}{No} & \textcolor{forestgreen}{No} & \textcolor{forestgreen}{No} \\
    \textbf{No License} & \textcolor{red}{High} & \textcolor{red}{High} & \textcolor{red}{High} & \textcolor{red}{High} & \textcolor{red}{High} & \textcolor{red}{High} \\
    \bottomrule
\end{tabular}}
\label{tab:license_reuse}
\end{table}

We use this rationale in \textbf{RQ2} to identify and categorize potential license noncompliance in our copy-based reuse network.
We use projects' latest status licenses for this examination.
Since both upstream and downstream projects may have multiple licenses, we evaluate all combinations of possible noncompliance to test hypothesis \textbf{H2a}.
However, there is an aggregation design decision here: how to aggregate possible noncompliance combinations of licenses \textit{with different risk levels} for the same pair of upstream--downstream projects?
We consider two options.
For a \textbf{high sensitivity} approach, we select the highest risk level combination of licenses for a given pair of upstream--downstream projects.
Conversely, for a \textbf{low sensitivity} approach, we select the lowest risk level combination.

\[
\text{Compliance}_{\text{A,B}} = 
\begin{cases} 
      \max \left( \text{risk}(L_{A_i}, L_{B_j}) \right): & \text{High Sen.} \\
      \min \left( \text{risk}(L_{A_i}, L_{B_j}) \right): & \text{Low Sen.} 
\end{cases}
\]

where:
\begin{itemize}
    \item \( L_{A_i} \): Each license of Project A,
    \item \( L_{B_j} \): Each license of Project B,
    \item \( \text{risk}(L_{A_i}, L_{B_j}) \): Incompatibility risk level between license \( L_{A_i} \) and license \( L_{B_j} \).
\end{itemize}

For brevity, we present and discuss only the low-sensitivity results below, but include the high-sensitivity results in our replication package, for completeness.

\subsection{Copy-based vs.\ Dependency-based Reuse}

To test our hypothesis \textbf{H2b}, we compare the reuse instances captured via copy-based network with dependency-based network.
For this analysis, we focus on high-risk categories in low-sensitivity mode with 10 or more reused blobs---our least conservative scenario---to quantify how often noncompliance is detectable through conventional methods (package manager analysis) versus cases that require copy-based detection.
To keep the analysis tractable we selected a sample of 50,000 unique upstream-downstream project pairs from our dataset.
Using a stratified sampling, we proportionally selected from each of the 16 high-risk categories, which together represent a total of 82 million projects.
To ensure adequate representation of smaller categories, a minimum sample size of 1,000 was enforced, even when the proportional size was smaller.
This approach ensures sufficient representation from smaller categories while maintaining overall proportionality.
Our final sample included a total of 57,341 project combinations.

Next, we used the maps provided in WoC, which detail all import and export statements in every blob for each commit.
By analyzing these maps, we identified all import/export statements within the projects in our sample\footnote{
Analyzed languages: Java, JavaScript, Python, R, Rust, Scala, C\#, Go, Groovy, Kotlin, and Perl.
}.
We then matched these statements between upstream and downstream projects to determine if they share any declared dependencies (i.e., the downstream project imports a package that the upstream project exports).

\subsection{Regression Model}\label{sec:reg}

In \textbf{RQ1}, we investigate whether the upstream project's license type affects the likelihood of its artifacts getting reused, testing hypotheses \textbf{H1a} and \textbf{H1b}.
Since the response variable is binary (1 if the project has introduced at least one reused blob, 0 otherwise), a logistic regression model is used.
It is the standard approach for binary outcomes and enables us to estimate the probability of reuse from various predictors~\cite{agresti2012categorical}.

\subsubsection{Stratified Sampling}

Given the scale and diversity of OSS projects, we employed a stratified sampling approach to ensure that our regression model accurately represents the OSS landscape~\cite{thompson2012sampling}.
Projects were divided into strata based on six key variables: number of commits, blobs, authors, forks, active months, and earliest commit time.
These variables reflect project size, activity, and history, all of which are likely to influence our outcome variables, as discussed in Sec.~\ref{sec:relwork} above.
The strata were defined as follows: number of commits (fewer than 500, 500–2000, and more than 2000), number of blobs (fewer than 10,000 and more than 10,000), number of authors (one author, 2–10 authors, and more than 10 authors), number of forks (no forks and at least one fork), and active months (fewer than three months and more than three months).
Additionally, we categorized projects into four historical eras based on their earliest commit time: the Foundational Era (before 1998), the Dot-com Boom and OSS Expansion (1998–2010), the Maturation and Mainstream Adoption phase (2010–2018), and the Modern Era with a Community Focus (2019–present).
This stratification resulted in 288 unique bins.
We sampled projects from each bin, yielding a final dataset of approximately half a million projects. While some bins contained fewer projects than anticipated due to uneven distribution, this approach ensures that our sample is representative of the broader OSS ecosystem, allowing for robust and generalizable conclusions from our analyses.

\subsubsection{Predictors}

Checking for correlations among predictors is crucial in regression models, as multicollinearity---strong correlations between predictors---can distort the results and reduce reliability~\cite{dormann2013collinearity}.
To manage multicollinearity, we applied a 0.6 correlation threshold.
Variables with correlations exceeding this threshold indicate overlapping information, and removing them helps mitigate multicollinearity while retaining the most important predictors and their portion of explained variance~\cite{vatcheva2016multicollinearity}.
The descriptive statistics for the remaining variables
are provided in Table~\ref{tbl:stat-l}.

\begin{table*}[ht]
\centering
\caption{Regression Model - Descriptive Statistics}
\resizebox{0.99\textwidth}{!}{
\begin{tabular}{llcccc}
  \toprule
  \textbf{Variable} & \textbf{Description} & & \textbf{Statistics} & & \\ 
  \midrule
  \midrule
  Reuse & Introduced at least 1 reused blob & Yes: & 444,144 (77.62\%) & No: & 128,029 (22.38\%) \\
  \midrule
   & & \textbf{5\%} & \textbf{Median} & \textbf{Mean} & \textbf{95\%} \\ 
  EarliestCommit & Time since the earliest commit & 05/08/2006 & 07/05/2017 & 01/31/2016 & 11/23/2021 \\
  LatestCommit & Time since the latest commit & 04/30/2011 & 02/17/2020 & 03/15/2019 & 04/28/2023 \\
  CoreAuthors & Authors with 80\%+ of commits & 1 & 2 & 8.62 & 17 \\
  Forks & Number of forks & 0 & 0 & 27.66 & 48 \\
  Commits & Number of commits & 2 & 155 & 2,982.63 & 5,770 \\
  Files & Number of files & 5 & 1,820 & 17,295.57 & 59,939.60 \\ 
  AdoptDelay & Earliest commit to license adoption (days) & 0 & 0 & 133 & 751 \\ 
  Burstiness &  (Latest - Earliest) / Active months & 0 & 1 & 1.87 & 6.37 \\
  \multicolumn{6}{c}{\dotfill} \\
  Language & \hspace{-0.2cm}JavaScript \hspace{1.15cm} C/C++ \hspace{1.1cm} Python & Java & PHP & Ruby & (Remaining) \\ 
  Counts (\%) & \hspace{-0.6cm}221,588 (38.72\%) \hspace{0.1cm} 82,551 (14.43\%) \hspace{0.1cm} 53,468 (9.34\%) & 50,372 (8.80\%) & 44,952 (7.86\%) & 18,592 (3.25\%) & 100,650 (17.59\%) \\ 
  \multicolumn{6}{c}{\dotfill} \\
  License & \hspace{0.0cm}No License \hspace{1.5cm} Permissive & \hspace{-1.3cm} Copyleft & Weak Copyleft & Conditional Open & Public Domain \\ 
  Counts (\%) & \hspace{-0.35cm} 263,974 (46.13\%) \hspace{0.7cm} 148,320 (25.92\%) & \hspace{-1.2cm} 60,925 (10.65\%) & 43,143 (7.54\%) & 30,933 (5.41\%) & 24,878 (4.35\%) \\
  \bottomrule
\end{tabular}}
\label{tbl:stat-l}
\end{table*}

While we removed highly correlated numerical variables to avoid multicollinearity, this approach cannot be directly applied to categorical variables.
Therefore, we included interaction terms between two categorical variables---license type and programming language---in our model to better capture the combined effect of these factors on reuse probability.
This approach allows us to account for potential interactions between these variables, offering a more nuanced understanding of how different license types may influence reuse within the context of specific programming languages.

Additionally, we applied sum contrasts for these two predictors, also known as effect coding, where each level of the predictor is compared to the overall mean of all levels.
This method allows for a more balanced interpretation of coefficient estimates, by contrasting each category with the overall mean rather than a specific reference category.
In sum contrasts, the coefficients for all levels, including the intercept, sum to zero, ensuring that one level’s coefficient is determined by the others, thereby maintaining balance and enhancing interpretability in the model.

\section{Results and Discussion}
\subsection{RQ1 - Regression Model}

\subsubsection{Our Findings}

To establish a baseline, we first modeled the probability of reuse based solely on the project's license type, without considering other potential factors.
This initial model showed a significant relationship between license type and reuse likelihood.
Specifically, projects with permissive, copyleft, or weak copyleft licenses were more likely to have their artifacts reused, while those with public domain licenses were less likely to be reused.

\begin{figure}[t]
    \centering
    \includegraphics[width=0.89\linewidth]{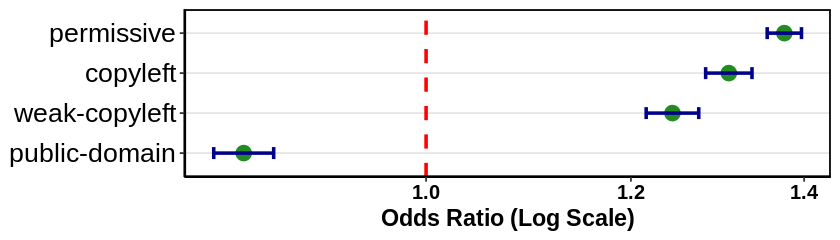}
    \caption{Simple Model - Odds Ratios and 95\% Confidence Intervals.}
    \label{fig:odds1}
\end{figure}

To assess the impact of the variables with significant coefficients, we examine the odds ratios derived from the logistic regression coefficients.
An odds ratio greater than 1 signifies a positive impact, whereas an odds ratio less than 1 indicates a negative impact.
Figure~\ref{fig:odds1} presents the odds ratios along with their corresponding 95\% confidence intervals.

Based on these findings, hypothesis \textbf{H1a} is partially supported.
Projects with permissive licenses, such as MIT and BSD, have higher reuse rates;
however, those with public domain licenses do not follow this pattern.
Similarly, hypothesis \textbf{H1b} receives partial support:
while restrictive licenses generally exhibit a lower probability of reuse compared to permissive licenses, they unexpectedly show higher odds of reuse than public domain licenses.

Recall, this initial model does not account for other potential factors that may influence reuse.
Consequently, while the preliminary results provide valuable insights, they may be confounded by unconsidered variables.
To address this limitation, we introduce a second model incorporating additional control variables, which allows for a more precise analysis of the true impact of license type on artifact reuse.

Table~\ref{tbl:anova-r} presents the ANOVA results for this model, showing that all predictors have highly statistically significant coefficients (p-values close to zero; not surprising given our sample size), and allowing for a comparison of relative explanatory power of each variable (the Deviance column).
Almost all control variables had the hypothesized effects, except for burstiness, which seems to be encouraging reuse; however, its deviance is relatively low.
The regression coefficient estimates are also shown in this table for non-categorical variables\footnote{
The p.value ($Pr(>|z|)$) for all this variables are close to zero ($<2.2e^{-16}$).
}.
Note, while a categorical variable may be significant in the model based on ANOVA results, indicating it contributes meaningfully, the coefficients for some individual \textit{levels} of the variable can still be insignificant.
This suggests that, although the variable as a whole impacts the outcome, not every category within it shows a statistically significant effect.

\begin{table}[t]
\centering
\caption{ANOVA Table and Regression Coefficients}
\begin{tabular}{lrrrr}
    \toprule
    & \textbf{Df} & \textbf{Deviance} &  \textbf{Pr($>$Chi)} & \textbf{Coefficient}\\ 
    \midrule
    EarliestCommit & 1 & 5,396 &  \cellcolor{lightgreen}$<2.2e^{-16}$ & $5.60e^{-01}$ \\ 
    LatestCommit & 1 & 30,987  & \cellcolor{lightgreen}$<2.2e^{-16}$ & $-1.49e^{-01}$ \\ 
    CoreAuthors & 1 & 8,143 &  \cellcolor{lightgreen}$<2.2e^{-16}$ & $2.56e^{-01}$ \\ 
    Forks & 1 & 7,994 &  \cellcolor{lightgreen}$<2.2e^{-16}$ & $4.05e^{-01}$ \\ 
    Commits & 1 & 23,749 & \cellcolor{lightgreen}$<2.2e^{-16}$ & $2.57e^{-01}$ \\ 
    Files & 1 & 65,912 &  \cellcolor{lightgreen}$<2.2e^{-16}$ & $2.80e^{-01}$ \\ 
    AdoptionDelay & 1 & 662 &  \cellcolor{lightgreen}$5.75e^{-146}$ & $2.05e^{-02}$ \\ 
    Burstiness & 1 & 128 &  \cellcolor{lightgreen}$1.42e^{-29}$ & $6.68e^{-02}$ \\ 
    Language & 11 & 7,710 & \cellcolor{lightgreen}$<2.2e^{-16}$ & Cat. \\ 
    License & 5 & 874 &  \cellcolor{lightgreen}$1.20e^{-186}$ & Cat. \\ 
    Language:License & 55 & 1,799 & \cellcolor{lightgreen}$<2.2e^{-16}$ & Cat. \\ 
    \bottomrule
\end{tabular}
\label{tbl:anova-r}
\end{table}

\begin{figure}[t]
    \centering
    \includegraphics[width=0.99\linewidth]{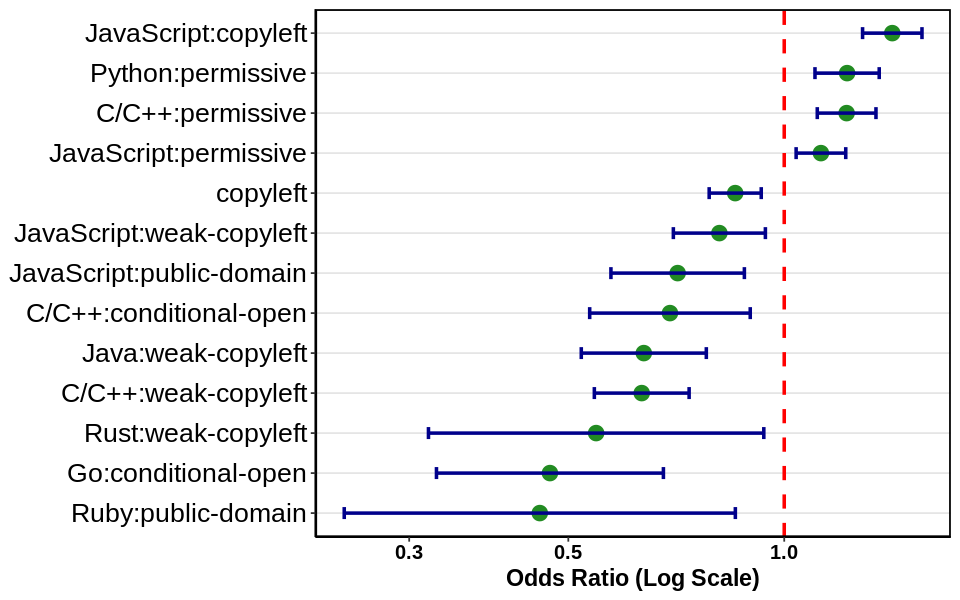}
    \caption{Full Model - Odds Ratios and 95\% Confidence Intervals.}
    \label{fig:odds2}
\end{figure}

Similarly to the previous model, Figure~\ref{fig:odds2} displays the odds ratios and their corresponding 95\% confidence intervals for the significant license variables.
When additional control variables such as programming language and its interaction with license types are introduced into the model, the results reveal a more nuanced understanding of how these license types influence software reuse.
Significant results are observed only in specific combinations of license types and programming languages.

For permissive licenses, Python, C/C++, and JavaScript projects exhibit an odds ratio greater than 1, indicating an increase in reuse.
The positive impact of permissive licenses is significant only for these three programming languages, while other languages do not show statistically significant effects.

The first model suggested that public domain licenses are negatively associated with reuse, and the second model confirms that this effect is significant only for JavaScript and Ruby, with no notable impact in other languages.
This finding implies that public domain licenses may lack the legal incentives or protections that developers value, making them less attractive for promoting reuse in certain contexts.
Hypothesis \textbf{H1a} is therefore partially supported.

Projects using permissive licenses show increased reuse in Python, C/C++, and JavaScript, but this effect is not significant in other languages, suggesting that permissive licenses enhance reuse only in specific environments.
Furthermore, public domain licenses do not generally impact reuse odds, but reduce the likelihood of reuse in JavaScript and Ruby, contrary to the expectation that more permissive licenses encourage reuse, and thus in contrast to \textbf{H1a}.

Several factors may contribute to this unexpected result.
One possibility is \textbf{legal uncertainty}; the concept of dedicating works to the public domain is not consistently recognized across jurisdictions.
In some countries, authors cannot fully waive their copyright, leading to ambiguities that might deter developers from reusing public domain code.
Additionally, the \textbf{absence of explicit permissions} can create confusion.
Although public domain status implies freedom of use, developers and organizations may prefer licenses that clearly state permissions and limitations, such as the MIT or BSD licenses, which provide explicit legal reassurances.
The perceived \textbf{lack of explicit disclaimers or warranties} in public domain software might also make it appear riskier, particularly for commercial use.
By contrast, permissive licenses typically include clauses limiting liability and disclaiming warranties, thereby offering additional protections.
\textbf{Community trust and familiarity} may also play a significant role.
Established permissive licenses are widely recognized and trusted, whereas public domain licenses may not enjoy the same level of familiarity or acceptance, leading developers to favor more well-known licensing options.

For copyleft licenses, the overall effect is negative.
However, JavaScript projects under such licenses exhibit an odds ratio greater than 1, suggesting that the effect of copyleft licenses varies significantly depending on the language.
Weak copyleft licenses also show negative impacts on reuse for JavaScript, Java, C/C++, and Rust projects.
These findings suggest that hypothesis \textbf{H1b} is also partially supported.
Although copyleft licenses generally reduce the probability of reuse, this is not the case for all programming languages.
Moreover, weak copyleft licenses reduce reuse only in specific languages.

\begin{figure}[t]
\small
\begin{tcolorbox}[colback=gray!5!white, colframe=gray!70!black, title=RQ1 Key Findings]
\begin{enumerate}[wide, labelwidth=!, labelindent=0pt]
    \item Permissive licenses have the strongest positive impact on reuse, particularly in Python, C/C++, and JavaScript projects. (H1a)
    \item Public domain licenses show a negative association with reuse, specifically in Ruby and JavaScript projects. (H1a)
    \item Copyleft licenses show mixed results: they are beneficial for reuse in specific contexts, such as JavaScript, but generally have a negative effect on reuse when controlling for other factors. (H1b)
    \item Weak copyleft licenses reduce reuse only in Rust, C/C++, and Java projects when other factors are considered. (H1b)
    \item The influence of license type on reuse is highly dependent on programming language, indicating that license effectiveness varies significantly across different language ecosystems.
\end{enumerate}
\end{tcolorbox}
\end{figure}

\subsubsection{Implications}

A key takeaway is that the choice of license for a project has a substantial impact on the likelihood of its artifacts being reused.
This effect varies across different license types and programming languages, highlighting nuanced relationships between license choice, programming language, and reuse behavior.
This indicates that developers and contributors should be mindful of how their choice of license can influence the adoption and reach of their work.

\looseness=-1
One of the most unexpected findings is that public domain licenses, designed to allow free and unrestricted reuse, have a negative effect on reuse.
This is concerning because the intent behind these licenses is to eliminate barriers, yet the data suggest the opposite.
The negative association of public domain licenses with reuse indicates that the OSS community may need to address this unintended outcome.
One way forward is to enhance awareness and education about public domain licensing, clarifying the legal protections and reuse rights it offers.
Clearer guidance on how public domain licenses differ from other open source licenses, particularly regarding legal clarity and potential liability, could benefit OSS contributors, especially newcomers.
The community might also consider providing stronger legal frameworks or support around public domain licenses to reduce uncertainties and hesitations.
Project maintainers may also reconsider using public domain licenses if their primary goal is to maximize reuse.
The data suggest that permissive licenses may be more effective in promoting reuse.

\looseness=-1
In conclusion, while the OSS movement encourages reuse and collaboration, these results show that the choice of license plays a crucial role in determining whether a project achieves those goals.
The community must be attentive to the barriers that certain licenses, such as public domain, may unintentionally create and take steps to provide better education, support, and legal frameworks to ensure that the intentions behind these licenses are effectively realized in practice.

\subsection{RQ2 - Noncompliance}

\subsubsection{Our Findings}

\begin{figure}[t]
    \centering
    \includegraphics[width=\linewidth, clip=true, trim=0 20 160 0]{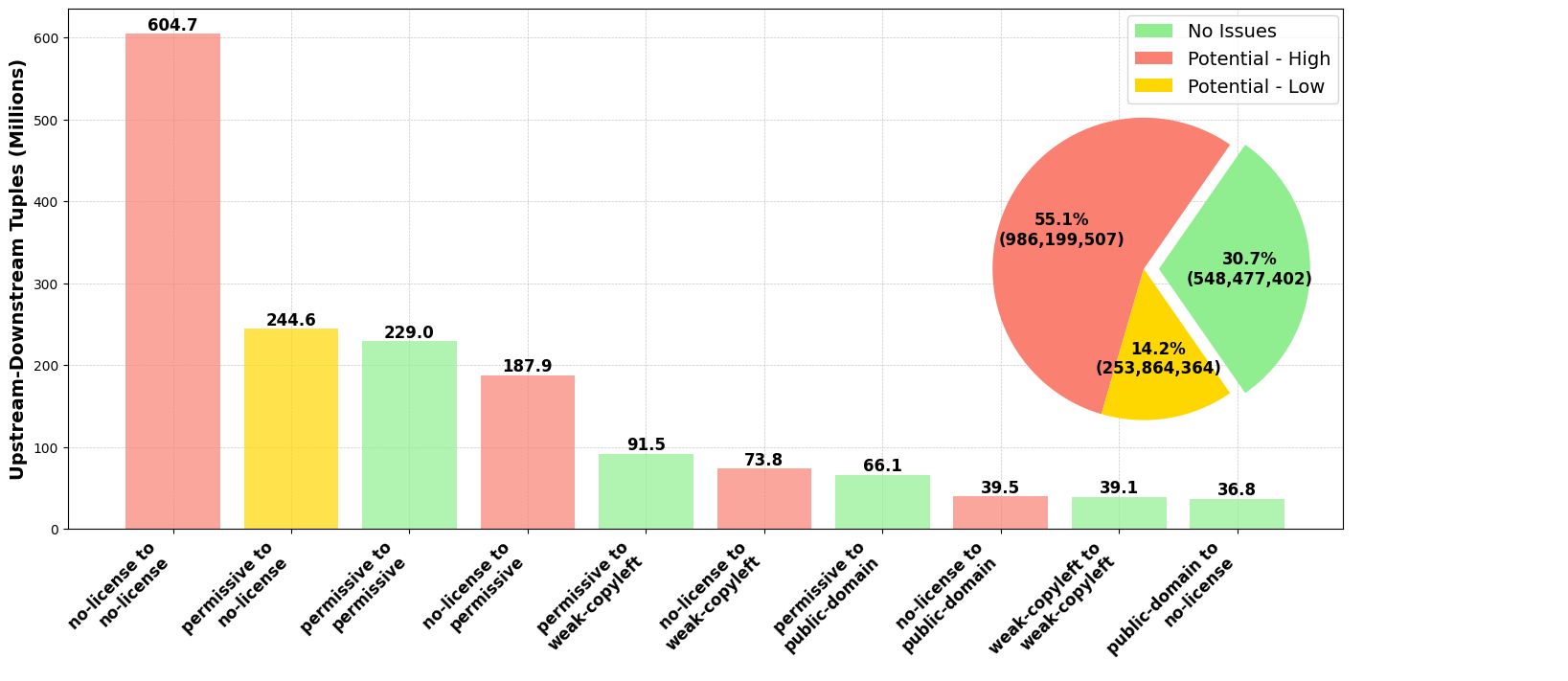}
    \caption{Top 10 License Types - 1 Reused Blob, Low Sensitivity}
    \label{fig:reuse0-l}
\end{figure}

\begin{figure}[t]
    \centering
    \includegraphics[width=\linewidth, clip=true, trim=0 20 160 0]{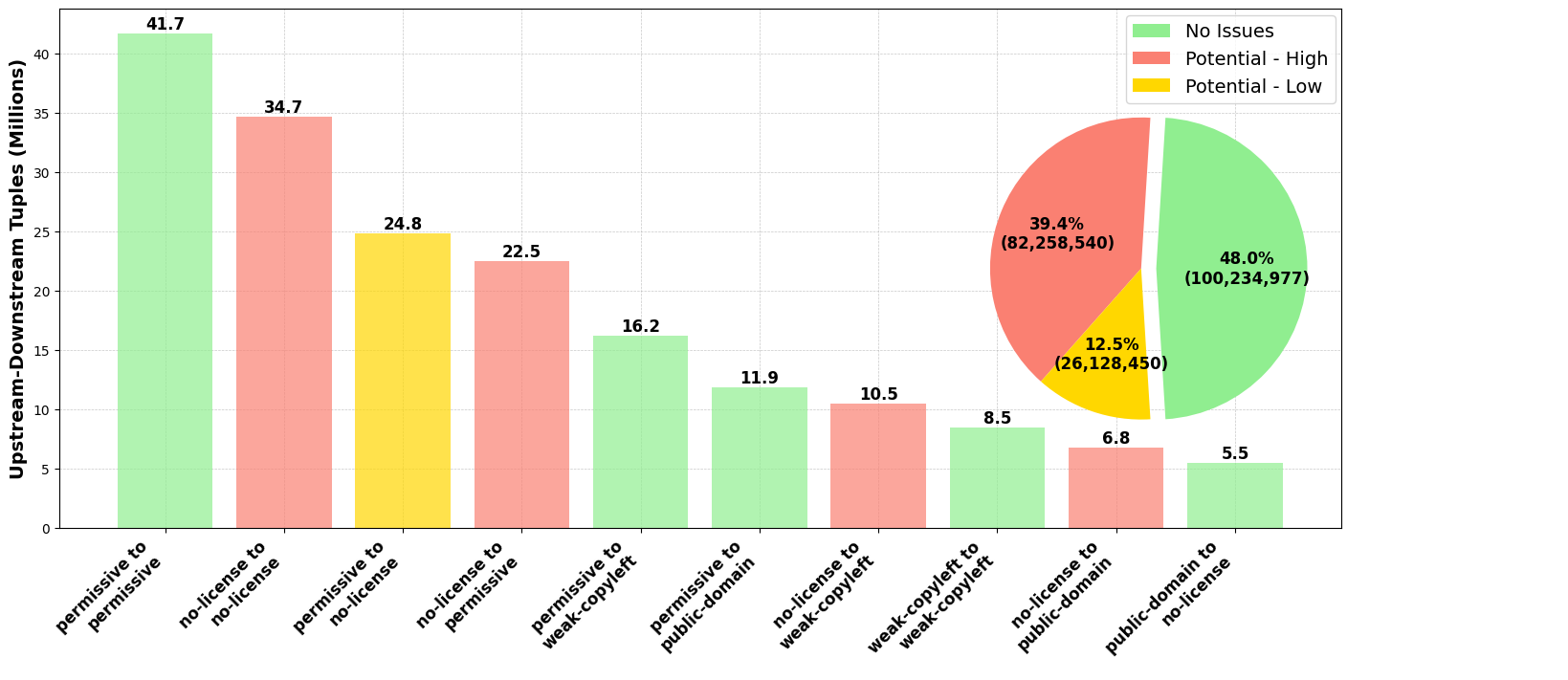}
    \caption{Top 10 License Types - 10 Reused Blobs, Low Sensitivity}
    \label{fig:reuse9-l}
\end{figure}

As discussed above, we report only the results of our low-sensitivity aggregation here (i.e., considering the lowest-risk pairs of licenses for a given upstream--downstream pair of projects).
Figures~\ref{fig:reuse0-l} and~\ref{fig:reuse9-l} summarize our findings for the two flavors of reuse we consider (complete reuse, with at least one shared blob, and substantial reuse, with 10 or more blobs).

\paragraph{At least 1 Reused blob}

Figure~\ref{fig:reuse0-l} highlights the top 10 categories of license combinations between upstream and downstream projects, showcasing the most frequent pairings.
The pie chart illustrates the distribution of project tuples across three categories: no issues, high-risk potential, and low-risk potential for license noncompliance.

\begin{figure}[t]
    \centering
    \includegraphics[width=\linewidth, clip=true, trim=0 20 160 0]{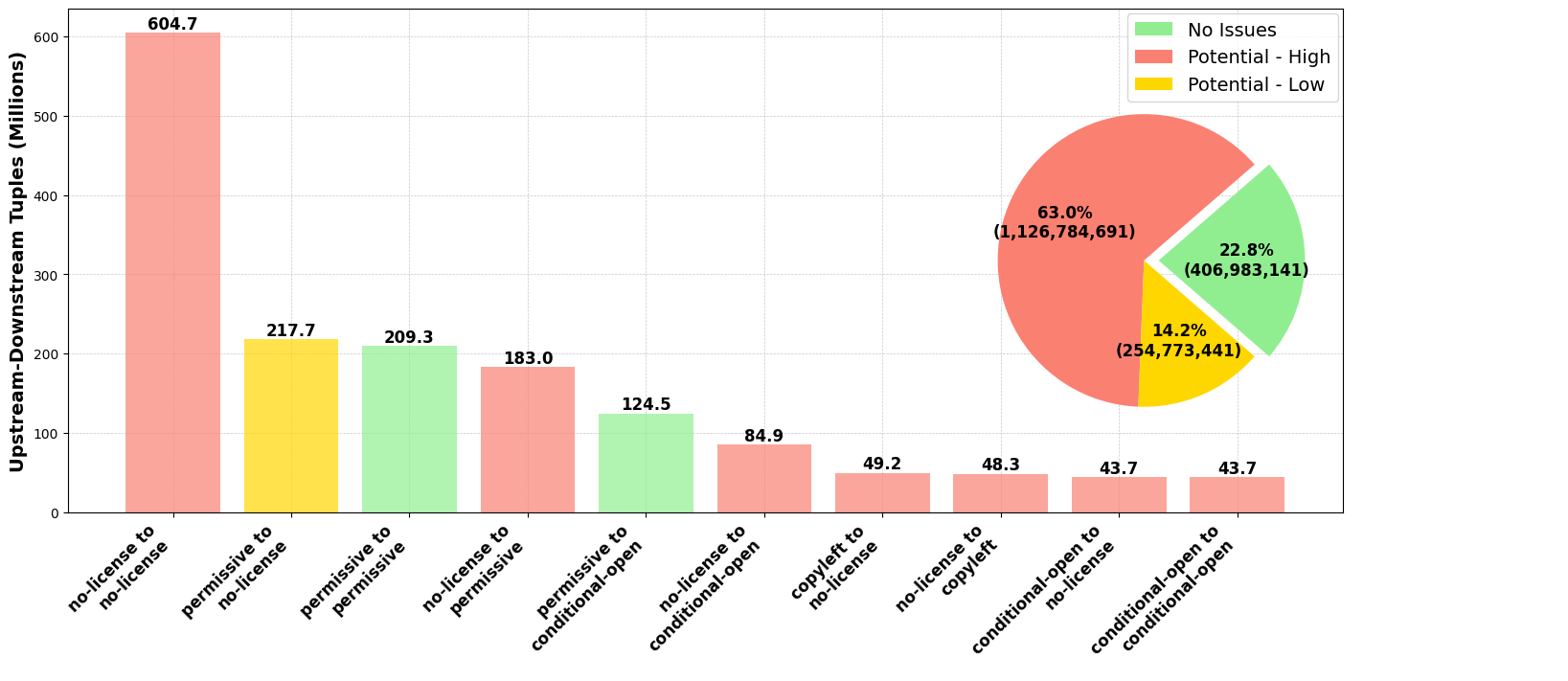}
    \caption{Top 10 License Types - 1 Reused Blob, High Sensitivity}
    \label{fig:reuse0-h}
\end{figure}

The results indicate that a significant majority (55\%) of upstream-downstream license combinations fall into the high-risk category.
The most common high-risk scenario occurs when neither the upstream nor downstream projects have a license, accounting for 605 million project tuples.
This creates legal uncertainty regarding reuse, modification, and distribution rights.
Other high-risk combinations within the top 10 involve cases where one project lacks a license, such as no-license to permissive.
Even when the upstream project has a clear license, the absence of a downstream license introduces ambiguity and potential legal challenges.

On the positive side, 30\% of the tuples present no issues, such as permissive to permissive combinations, where both upstream and downstream projects are clearly licensed, minimizing legal risk.
Low-risk combinations make up 14\%, including cases like permissive to no-license, which involves some legal uncertainty but is less risky than high-risk scenarios.

As Figure~\ref{fig:reuse0-l} shows, the proportion of high-risk tuples decreases from 63\% to 55\%, with many of these tuples shifting to the no-issues category, which increases from 23\% to 31\%.
This shift is primarily observed in the permissive to weak-copyleft and public-domain combinations, indicating that permissive licenses are prevalent in projects with multiple licenses.
Despite these shifts, the no-license to no-license combination remains the largest high-risk group, unchanged, highlighting the ongoing legal uncertainty in projects without clear licensing.

\paragraph{At least 10 Reused blobs}

The total number of reuse instances (unique combinations of upstream and downstream projects) drops significantly from 1.816 billion to 212 million after applying the constraint of at least 10 reused blobs—a reduction of approximately 88\%.
This sharp decline indicates that the majority of earlier reuse cases involved fewer than 10 blobs, suggesting that much of the initial reuse was minimal or partial.
This reduction highlights that a significant portion of copy-based reuse in the open source ecosystem is small-scale or potentially superficial, involving limited sharing between projects, with fewer instances of deeper, substantial dependencies.
By focusing on reuse instances involving at least 10 reused blobs, the data now captures more meaningful relationships, wherein downstream projects are more closely integrated with upstream codebases.

Although the number of high-risk combinations decreases proportionally from the earlier results, they still account for 39\% of the remaining reuse instances (see Figure~\ref{fig:reuse9-l}).
This indicates that even in cases of more substantial reuse, issues related to licensing or lack of clear licensing persist.
However, the increase in the no-issues category to 48\%, primarily driven by permissive to permissive license reuse, suggests that when significant reuse occurs, clearer licensing tends to be in place, especially for permissive licenses.

\begin{figure}[t]
    \centering
    \includegraphics[width=\linewidth, clip=true, trim=0 20 160 0]{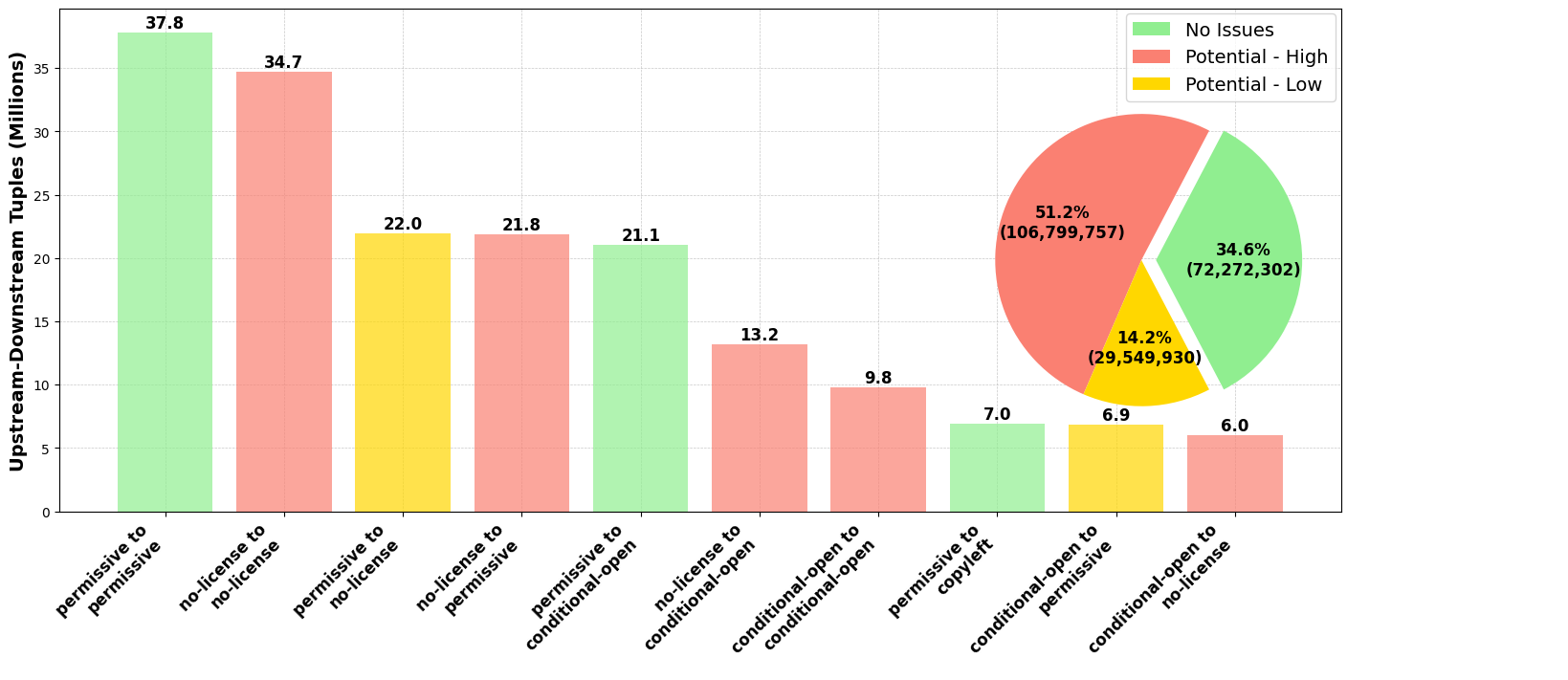}
    \caption{Top 10 License Types - 10 Reused Blobs, High Sensitivity}
    \label{fig:reuse9-h}
\end{figure}

Under the low-sensitivity approach for reuse involving over 10 blobs, there is a noticeable shift in the distribution of risk categories (see Figure~\ref{fig:reuse9-l}).
High-risk combinations decrease from 51\% in the high-sensitivity scenario to 39\%, indicating a more favorable risk landscape when the least-risky license is selected in multi-licensed projects.

The no-issues category grows to 48\%;
however, the no-license to no-license combination remains a significant high-risk category, unchanged at 35 million tuples.
This persistent issue demonstrates that, even with more lenient interpretations, projects lacking clear licenses continue to pose substantial legal risks.

Overall, while the results under the low-sensitivity approach reveal a more favorable risk profile, with fewer high-risk combinations and an increased share of safe reuse, we still observe 82 million reuse instances (39\% of total reuse) with high potential risk of license noncompliance, supporting our Hypothesis H2a and underscoring the critical importance of proper licensing.

Overall, these findings support our hypothesis \textbf{H2a} and underscore the critical importance of proper licensing.

\paragraph{Copy-based vs.\ Dependency-based Reuse}

Existing research highlights the need for more granular methods to detect code reuse beyond declared dependencies, as traditional tools often overlook directly copied code between projects.
This section aims to expose the limitations of relying solely on package managers for license compliance monitoring.
While package managers effectively track formal dependencies (such as imports and exports) between projects, they often fail to detect instances of direct code copying without an explicitly declared dependency.

The results of comparing reuse detected via copy-based network and dependency-based network are presented in Table~\ref{tbl:depend-comp}.

\begin{table}[t]
\centering
\caption{Reuse Detectable by Dependency Relationship}
\resizebox{0.99\linewidth}{!}{
\begin{tabular}{lrrr}
    \toprule
    \textbf{License Type} & \textbf{Sample Size} & \textbf{Decl.\ Dep.} & \textbf{Percent} \\ 
    \midrule
    no-license-2-no-license & 21,102 & 499 & 2.36\% \\
    no-license-2-permissive & 13,670 & 346 & 2.53\% \\
    no-license-2-weak-copyleft & 6,357 & 94 & 1.48\% \\
    no-license-2-public-domain & 4,107 & 93 & 2.26\% \\
    copyleft-2-no-license & 1,105 & 48 & 3.35\% \\
    conditional-open-2-conditional-open & 1,000 & 20 & 2.00\% \\
    conditional-open-2-copyleft & 1,000 & 20 & 2.00\% \\
    conditional-open-2-no-license & 1,000 & 40 & 4.00\% \\
    conditional-open-2-weak-copyleft & 1,000 & 18 & 1.80\% \\
    copyleft-2-conditional-open & 1,000 & 19 & 1.90\% \\
    copyleft-2-permissive & 1,000 & 36 & 3.60\% \\
    copyleft-2-public-domain & 1,000 & 36 & 3.60\% \\
    copyleft-2-weak-copyleft & 1,000 & 14 & 1.40\% \\
    no-license-2-conditional-open & 1,000 & 34 & 3.40\% \\
    no-license-2-copyleft & 1,000 & 43 & 4.30\% \\
    weak-copyleft-2-no-license & 1,000 & 36 & 3.60\% \\
    \midrule
    \textbf{Total} & \textbf{57,341} & \textbf{1,396} & \textbf{2.43\%} \\
    \bottomrule
\end{tabular}}
\label{tbl:depend-comp}
\end{table}

The results highlight a significant limitation in current dependency detection tools, showing that the percentage of code reuse detected through declared dependencies is remarkably low across all categories.
Despite analyzing over 57,000 project combinations, the overall detection rate of code reuse through formal dependency relationships was only 2.43\%.
This suggests that traditional methods relying on package managers, which track declared imports and exports between projects, are insufficient for capturing most instances of code reuse, supporting our hypothesis \textbf{H2b}.

\begin{figure}[t]
\small
\begin{tcolorbox}[colback=gray!5!white, colframe=gray!70!black, title=RQ2 Key Findings]
\begin{enumerate}[wide, labelwidth=!, labelindent=0pt]
    \item A significant portion of upstream-downstream project combinations are classified as high-risk for potential license noncompliance, leading to considerable legal uncertainties regarding reuse, modification, and distribution rights. (H2a)
    \item The most common high-risk potential noncompliance scenario involves projects lacking any license, underscoring a legal vulnerability within the open-source community and highlighting the urgent need for consistent and clear licensing practices.
    \item Dependency tracking is inadequate for detecting most instances of code reuse, highlighting the need for more granular detection methods capable of identifying copy-based reuse that would enable more accurate license compliance monitoring in open-source projects. (H2b)
\end{enumerate}
\end{tcolorbox}
\end{figure}

\subsubsection{Implications}

These findings have significant implications for the open-source community, particularly in relation to license compliance and code reuse detection.
The results reveal that a majority of upstream-downstream project combinations are classified as high-risk for potential noncompliance, underscoring a persistent issue in open-source software development.
The high occurrence of high-risk cases, especially in projects with no license, highlights a potential legal vulnerability that could impact the sustainability and collaboration within the open-source ecosystem.

This findings also call for more advanced detection techniques that go beyond traditional dependency analysis.
Tools that can detect code reuse through copying
are essential for identifying non-compliance with licensing terms.
The low detection rates across the board demonstrate that current tools are not capable of providing a complete picture of how code is reused, and more comprehensive approaches are necessary to ensure effective license compliance monitoring.

\section{Limitations}
\subsection{Internal Validity}

\subsubsection{Project to License Map} 

\looseness=-1
The project to license map (P2L) in WoC relies on detecting license files in repositories, assuming licenses are always recorded in dedicated files.
Nevertheless, licenses might appear in README or source files, leading to underreporting or misclassification.
This suggests that results should be interpreted cautiously, and additional manual verification may be needed for a more accurate understanding of license noncompliance.

\subsubsection{License Scope}

Assigning a license to an entire OSS project can introduce challenges, as the license may not uniformly apply to all components.
Projects often incorporate third-party libraries, modules, or contributions that come with their own distinct licenses, which may conflict with or restrict the applicability of the main project license.
Thus, while the project may be licensed under a specific open-source framework, that license may only cover certain parts, with other components subject to different licensing terms.

\subsubsection{Dependency-Based Reuse}

\looseness=-1
One limitation in comparing copy-based reuse with dependency-based reuse is that some projects use dynamic or implicit imports, where dependencies are loaded at runtime or through unconventional methods that may not be captured by a straightforward export-import analysis.
This can result in certain dependencies, which package managers can detect, being overlooked, exposing gaps in our approach.
Nonetheless, our methodology is conservative, as we track dependencies over time rather than focusing solely on the latest version.
By excluding any reuse instance that was detectable through dependencies at any point in the project's history, we provide a more thorough view of potential dependency-based reuse. 
This approach reduces the risk of missing past dependencies that may have been removed or modified in subsequent versions, delivering a more inclusive analysis of reuse instances.
However, this conservatism may also lead to attributing reuse to dependencies that no longer exist, slightly skewing the results toward historical dependency detection.

\subsection{External Validity}

\subsubsection{Copy-Based Reuse}

While emphasizing copy-based reuse offers valuable insights into license compliance, we recognize the significant role of dependency-based reuse within the broader reuse network.
Focusing solely on copy-based reuse may overlook certain aspects of how dependencies are integrated into a project.
Conversely, dependency-based reuse can miss critical instances where code is directly copied between projects, which is equally crucial in identifying potential noncompliance.
Thus, while this work prioritizes copy-based reuse, it serves to complement—rather than replace—the understanding gained from analyzing dependency-based reuse, together providing a more comprehensive view of compliance.

\section{Conclusions}

Our study shows that the choice of open-source license plays a significant role in influencing the likelihood of reuse.
Permissive licenses consistently encourage reuse across a variety of programming languages, while copyleft and weak copyleft licenses exhibit more context-specific effects, sometimes limiting reuse depending on the language and environment. 
Despite offering unrestricted reuse, public domain licenses were linked to a negative impact on reuse, likely due to legal uncertainties.
Our findings also emphasize the importance of detecting copy-based reuse, as traditional dependency-based approaches often fail to capture the full scope of reuse, especially when explicit dependencies are not declared.
This highlights the need for more advanced detection methods to improve license compliance monitoring in the open-source ecosystem.
Moreover, projects without clear licenses continue to present significant legal risks, underscoring the need for more consistent and transparent licensing practices within the open-source community.

\clearpage
\balance
\printbibliography

\clearpage
\appendix
\section{License Types}\label{ap:type}
List of SPDX license identifiers aggregated by their respective license types:

\textbf{Permissive}: 0BSD, AFL-3.0, Apache-2.0, BSD-2, BSD-2-Clause, BSD-3-Clause, BSL-1.0, ISC, Libpng, MIT, MIT-0, MITNFA, MIT-Wu, MS-PL, OpenSSL, PHP-3.01, Pixar, PSF-2.0, Ruby, SGI-B-2.0, TCL, WTFPL, Zlib

\textbf{Copyleft}: deprecated\_AGPL-3.0, deprecated\_GPL-3.0+, GPL-2.0, GPL-3.0+, GPL-CC-1.0, OSL-3.0

\textbf{Weak Copyleft}: Artistic-1.0-Perl, Artistic-2.0, CDDL-1.0, deprecated\_LGPL-2.1, eprecated\_LGPL-3.0, EPL-1.0, EPL-2.0, LGPL-2.0+, LGPL-3.0, MPL-1.1, MPL-2.0-no-copyleft-exception

\textbf{Conditional Open}: CC-BY-3.0, CC-BY-4.0, CC-BY-SA-3.0, CC-BY-SA-4.0, ODC-By-1.0, OFL-1.0, OFL-1.1

\textbf{Public Domain}: CC0-1.0, libtiff, Unlicense

\end{document}